\documentclass[dvipdfmx]{elsarticle}
\usepackage{lineno}
\modulolinenumbers[5]

\journal{Magnetic Resonance Imaging}

\usepackage{amssymb}

\begin{document}

\newcommand{\Rset}{\mathbb{R}}
\newcommand{\InFSE}{InFS--Explorer$^{\tiny{\copyright{}}}$}
\newcommand{\InFSEkisacik}{InFSE$^{\tiny{\copyright{}}}$}
\newcommand{\InFSEkisa}{InFSExp$^{\tiny{\copyright{}}}$}

\def\dblone{\hbox{$1\hskip -1.5pt\vrule depth 0pt height 1.6ex width 1pt
  \vrule depth 0pt height 1pt width 0.1em$}}

\begin{frontmatter}

\title{
On Using Signal Magnitude in Diffusion Magnetic Resonance Measurements
of Restricted Motion
}

\author[acuaddress]{Alpay \"Ozcan}
\ead{alpay.ozcan@acibadem.edu.tr}

\address[acuaddress]{Biomedical Imaging Research and Development Laboratory, and
Department of Biomedical Device Technologies,
Ac{\i}badem Mehmet Ali Ayd{\i}nlar University, Kay\i\c{s}da\={g}{\i} Cad.,
No:32 34752, Ata\c{s}ehir, Istanbul, Turkey}

\begin{abstract}
Tissue microstructure has significance as a biomarker,
however its accurate inference with diffusion magnetic resonance (MR)
is still an open problem.
With few exceptions, diffusion weighted (DW) MR models
either process diffusion MR data using signal magnitude,
whereby microstructural information is forcefully confined to symmetry
due to Fourier transform properties, or directly use symmetric basis expansions.

Herein, information loss from magnitude utilization is demonstrated
by numerically simulating particles undergoing diffusion
near a fully reflective infinite wall and an orthogonal corner.

Simulation results show that
the loss of the Hermitian property when using signal magnitude impedes
DW--MR from accurately inferring microstructural information in both of the geometries.


\end{abstract}

\end{frontmatter}


\pagestyle{myheadings}
\markboth{\"Ozcan, June, 23, 2019}
{\"Ozcan, June, 23, 2019}

\date{June, 23, 2019}

\section{Introduction}
Tissue microstructure, which can be inferred by molecular
motion measurements, has high significance as a biomarker.
However, accurate inference of restricted motion
with diffusion MR is still an open problem.
Several models are in use~\cite{alpaybackground12},
and with few exceptions~\cite{ozarslan201316,moseley04gdt},
DW--MR models
process diffusion MR data using signal magnitude
thereby confining microstructural information
to symmetry due to Fourier transform properties.

Symmetry assumptions might be realistic in isotropic environments such
as homogeneous liquids.
By contrast, in a complex environment,
such as biological tissue, the path traveled by the molecules
is determined by the
heterogeneous microstructure and thereby
displacement asymmetry becomes possible.
Two existing basis expansion models, the mean apparent
propagator (MAP) model~\cite{ozarslan201316} and
generalized diffusion tensor~\cite{moseley04gdt} model,
recognize asymmetry and work on the complex valued DW--MR signal.
Likewise, the more recently developed complete Fourier direct magnetic
resonance imaging (CFD--MRI)~\cite{alpaycfd} methodology
treats the complex valued signal and the distribution of the
displacement integrals without any symmetry constraints
by establishing a high dimensional Fourier relationship
between them.

However, direct utilization of symmetric templates
in model matching methods such as symmetric tensors of diffusion
tensor imaging (DTI)~\cite{basser94,matiello94}
or spherical harmonics of high angular resolution
diffusion imaging (HARDI)~\cite{frank01},
and/or symmetry by way of signal magnitude in the Fourier based
$q$--space methodologies~\cite{callaghan2011,wedeen05,tuch04} is now standard.
This raises the concern that information is being automatically
distorted and/or lost by collapsing the asymmetry within the signal,
thereby impeding accurate inference.
As analytical solutions of DW--MR signal processing exist
only for simple geometries such as infinite parallel plates~\cite{Cotts1966264,Robertson1966273},
planar, cylindrical and spherical
geometries~\cite{BARZYKIN1999342},~\cite{callaghan2011};
a proof of principle is needed for emphasizing and revising the importance
of concerns regarding loss of information when using symmetry based methods.

Herein, DW--MR inference inaccuracies when using signal magnitude
are demonstrated by simulating
MR signal from
diffusing molecules near
an infinite reflective wall and an orthogonal corner.
While these two basic
geometries
are not necessarily representatives of biological tissue's microstructure,
they provide solid test beds for analyzing and comparing
complex valued data versus magnitude usage in
the treatment of diffusion MR signal originating from asymmetric environments.

For properly demonstrating the fundamental concept,
the simulations were comprised of pulsed gradient spin
echo (PGSE) nuclear magnetic resonance (NMR) experiments
without any imaging gradients.

\section{Material and Methods}
\subsection{Theory}
\label{theorysection}
\begin{figure}[h!]
\centering
\includegraphics[width=12cm]{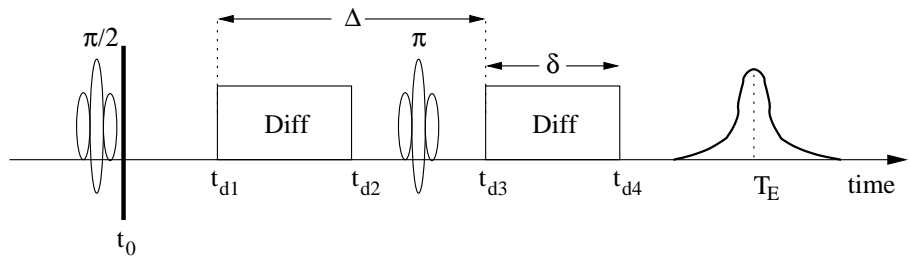}
\caption{The PGSE--NMR pulse sequence and the definition of the variables
used in the calculations.
Sampling starts before the echo time $T_E$ to capture the peak value of
the spin echo which is attenuated according to the motion sensitizing magnetic
field gradients.}
\label{pgseNMRfig}
\end{figure}%
For modeling coherent or incoherent motion,
the time--dependent position of
magnetic moments is represented in the most general fashion~\cite{alpaycfd}:
\begin{equation}
x_i(t)=x_i(t_0)+w_i(t)
\label{motion}
\end{equation}
using non--interacting particles.
In Eq.~(\ref{motion}), $w_i(t)\in \Rset^3$ represents the displacement of the
$i^\mathrm{th}$ magnetic moment from its initial position $x_i(t_0)$  (i.e., $w_i(t_0)=0$).
The initial time $t_0$ is chosen as the end time of
the $\displaystyle{\pi/2}$ radio frequency (RF) pulse (see Fig.~\ref{pgseNMRfig}).
The model covers any type of motion with only the assumption
of path continuity since a magnetic moment
cannot disappear at a given point and reappear at another.

The NMR signal, on the other hand, is modeled as the sum of the
individual transverse magnetization vectors~\cite{haacke},
$m_i,$ which is written in complex number form:
\begin{equation}
M(t)=\sum\limits_i m_i(t)=
m_0\,\sum\limits_i e^{-\jmath\,\gamma\, \Omega_i}.
\label{sumofpackets}
\end{equation}

Here, $\gamma$ is the gyromagnetic ratio and
$m_0$ is the initial magnetization tipped to the transverse plane at the end
of the $\displaystyle{\pi/2}$
RF pulse at $t_0$ (see Fig.~\ref{pgseNMRfig}).
The phase is obtained by multiplying $\gamma$ and $\Omega_i,$ which
is ideally a function of the magnetic field gradients
and the position of the magnetic moment $x_i\in \Rset^3.$
When ideal rectangular gradient pulses with amplitude vector
$G_\mathrm{D} \in \Rset^3$  as in Fig.~\ref{pgseNMRfig}
are applied, $\Omega_i$ becomes:
\begin{equation}
\Omega_i
=G_\mathrm{D} \cdot \left(
\int\limits_{t_{d3}}^{t_{d4}}x_i(\tau)\, d\tau
-\int\limits_{t_{d1}}^{t_{d2}}\,x_i(\tau)\, d\tau
\right).
\label{omegadefinition}
\end{equation}
where $t_{d1},$ $t_{d3}$ represent on-- and $t_{d2},$ $t_{d4}$
off--times of the
pulsed gradients.

Following the derivations in~\cite{alpaycfd}
that use the time parameters of Fig.~\ref{pgseNMRfig},
the DW--MR signal model's central theme is the displacement integral.
For the
PGSE sequence of Fig.~\ref{pgseNMRfig}
the displacement integral of the $i^\mathrm{th}$ magnetic moment is defined as~\cite{alpaycfd}:
\begin{equation}
W^\mathrm{d}_i \doteq \int\limits_{t_{d3}}^{t_{d4}}x_i(\tau)\, d\tau
-\int\limits_{t_{d1}}^{t_{d2}}\,x_i(\tau)\, d\tau
=\int\limits_{t_{d3}}^{t_{d4}}w_i(\tau)\, d\tau
-\int\limits_{t_{d1}}^{t_{d2}}\,w_i(\tau)\, d\tau \in \Rset^3.
\label{stochasticphase1}
\end{equation}
Additionally, the PGSE parameters $\delta$ (pulse length) and $\Delta$ (pulse separation)
are defined as:
$\delta \doteq t_{d2}-t_{d1}=t_{d4}-t_{d3}$ and
$\Delta \doteq t_{d3}-t_{d1}=t_{d4}-t_{d2}$ (see also Fig.~\ref{pgseNMRfig}).
This is basically the difference of the integrals of the paths
traveled by magnetic moments during the pulsed gradients.

According to Eq.~(\ref{sumofpackets})
and applying Eq.~(\ref{omegadefinition}) for the pulsed gradients of the PGSE sequence of Fig.~\ref{pgseNMRfig},
the total magnetization at the echo time $T_E$ is a function of
the pulsed gradient vector (i.e.\ the motion sensitizing magnetic field gradient vector),
$G_\mathrm{D},$ and the displacement integral $W^\mathrm{d}_i$
\begin{equation}
S_{\mathrm{cfd}}^{\mathrm{nmr}}=m_0\,\sum\limits_i e^{-\jmath\,\gamma\, G_\mathrm{D} \cdot W^\mathrm{d}_i}.
\label{NMRparticlesum}
\end{equation}

Rather than computing Eq.~(\ref{NMRparticlesum}) over the (finite) count of magnetic moments,
a more convenient expression is obtained by defining
$P_{\mathrm{cfd}}^{\mathrm{nmr}}(W^\mathrm{d})$ as the fraction of magnetic moments with
displacement integral value equal to $W^\mathrm{d}$ (per unit interval, $dW^d$).
Using the distribution of displacement integrals,
$P_{\mathrm{cfd}}^{\mathrm{nmr}}(W^\mathrm{d}),$
$S_{\mathrm{cfd}}^{\mathrm{nmr}}$ at the echo time $T_E$ in Fig.~\ref{pgseNMRfig}
is calculated in the (continuous) displacement integral space
for obtaining the signal $S_{\mathrm{cfd}}^{\mathrm{nmr}}:$
\begin{equation}
S_{\mathrm{cfd}}^{\mathrm{nmr}}
=\int P_{\mathrm{cfd}}^{\mathrm{nmr}}(W^\mathrm{d})\, e^{-\jmath\,\gamma\, G_\mathrm{D} \cdot W^\mathrm{d}}\, dW^\mathrm{d},
\label{cfdNMRresult}
\end{equation}
where both sides of the equation are divided by $m_0$
for ease of notation.
This formulation
is general and existing models can in fact be derived starting from
Eq.~(\ref{cfdNMRresult})~\cite{alpayisbi11,alpaybackground12}.

The expression of Eq.~(\ref{cfdNMRresult}) is exactly the Fourier
transform of
$P_{\mathrm{cfd}}^{\mathrm{nmr}}$ evaluated at the three
dimensional frequency vector $k_\mathrm{D}:$
\begin{equation}
S_{\mathrm{cfd}}^{\mathrm{nmr}}(k_\mathrm{D})=\mathcal{F}\{P_{\mathrm{cfd}}^{\mathrm{nmr}}\}(k_\mathrm{D}),
\label{CFDNMRsignal}
\end{equation}
where $k_\mathrm{D}=[k_{\mathrm{D}1},\, k_{\mathrm{D}2},\, k_{\mathrm{D}3}]=\gamma\, G_\mathrm{D}
\in \Rset^3$ represents the pulsed (or diffusion/motion sensitizing)
gradient vector with its entries equal to the amplitudes of the pulsed magnetic
field gradients of PGSE~\cite{alpaycfd} scaled by $\gamma.$

In other words, for each point $k_\mathrm{D} \in \Rset^3,$
the Fourier transform of $P_{\mathrm{cfd}}^{\mathrm{nmr}}$
is sampled at the corresponding acquisition.
Consequently,
$P_{\mathrm{cfd}}^{\mathrm{nmr}},$ which is the physical quantity
of interest, is recovered with the inverse Fourier transform
of the signal obtained from the MR scanner:
\begin{equation}
P_{\mathrm{cfd}}^{\mathrm{nmr}}(W^\mathrm{d})=\mathcal{F}^{-1}\{S_{\mathrm{cfd}}^{\mathrm{nmr}}\}(W^\mathrm{d}).
\label{CFDDistributionFourierinverse}
\end{equation}

Using the properties of the Fourier transform, Eq.~(\ref{CFDDistributionFourierinverse})
provides crucial information on the nature of
$S_{\mathrm{cfd}}^{\mathrm{nmr}}.$
The Fourier transform of a real valued function is a
{\bf complex valued} Hermitian function~\cite{bracewell}.
By the reciprocity property of
the Fourier transform~\cite{bracewell}
only symmetric real valued functions' Fourier transforms are real valued.
Therefore,
$S_{\mathrm{cfd}}^{\mathrm{nmr}}(k_\mathrm{D})$
in Eq.~(\ref{CFDNMRsignal}) is {\bf complex valued} Hermitian
since $P_{\mathrm{cfd}}^{\mathrm{nmr}}(W^\mathrm{d})$ is real valued.
Furthermore, when there are coherent displacements such as bulk motion,
Eq.~(\ref{stochasticphase1}) incorporates a bias term which in turn
reflects as a phase shift in
$S_{\mathrm{cfd}}^{\mathrm{nmr}}(k_\mathrm{D})$~\cite{alpaycfd}.

By contrast, existing $q$--space models~\cite{callaghan2011,wedeen05,tuch04}
use the magnitude of the signal in practice
when evaluating the Fourier transform with the hope of
filtering out bulk motion~\cite[pp.1378]{wedeen05}.
However,
experimental DW--MR signal is not necessarily Hermitian
symmetric as various factors including but not limited to susceptibility
distort it, whereby the signal magnitude becomes asymmetric.
When Fourier transformed,
the transform of the asymmetric signal magnitude is a
{\bf complex valued} Hermitian function
and therefore cannot describe a physically meaningful distribution.
The solution offered by the $q$--space methods is to take the magnitude one more time
creating a real symmetric function due to Hermitian symmetry in the complex domain.

In short, by taking the magnitude before and after the Fourier transform,
$q$--space methods constrain the outcome to be real valued and symmetric.
This raises the concern that more than what was initially intended
(e.g.,\ bulk motion removal) is in fact being eliminated unnecessarily.

As it is impossible to fully analytically describe the distribution
functions of the displacement integrals,
herein, simulations were used to obtain the functions
numerically for showing the detrimental effects of using signal
magnitude in the calculations.

\subsection{Simulations}
\label{simulationssection}
Scenarios imitating water protons near reflective walls were
constructed with $n_\mathrm{particles}=120000$ particles.
With a fixed simulation step size (the standard deviation of the increments),
non--interacting particle motion was implemented using a hindered random walk with
normally distributed increments ($w_i(t)$ in Eq.~(\ref{motion})) with zero mean.
This procedure simulated molecular motion propelled by thermal energy
(see~\cite{Ermak19754189} for an early simulation example).
For modeling hinderance caused by the structure, the displacements were modified
when a path crossed a wall.
Effectively, the path was readjusted by computing the outcome of the elastic collision(s).

There were $3000$ simulation steps executed for each particle.
For matching the experimental values of PGSE parameters used in an earlier
biological phantom work~\cite{alpayembc11}
$(\delta=15\mbox{ms}$ and $\Delta=30\mbox{ms}),$
simulation values of the same parameters
were chosen as $\delta_\mathrm{sim}=1000$ steps and $\Delta_\mathrm{sim}=2000$ steps
with a step time of
$\displaystyle{t_{\mathrm{step}}=15 \times 10^{-6}\,\mbox{s}},$
e.g.\ $\delta=t_{\mathrm{step}}\,\delta_\mathrm{sim}.$

The step size, $s,$ of the random walk was calculated based on the
diffusion coefficient of water, $D_{\mathrm{water}}=2.6\times 10^{-3}\,\mbox{mm}^2/\mbox{s}:$
\begin{equation}
s=\sqrt{6\,D_{\mathrm{water}}\, t_{\mathrm{step}}}=0.48\, \mu\mbox{m}.
\label{stepsizeeq}
\end{equation}
Numerical simulations consisted of:\\
\noindent{\sf DW--MR Simulations}
\begin{enumerate}
\label{numericalsimsteps}
\item At the initial time step, $k=0,$ uniformly distributing initial
positions of the particles near the walls,
\item
\label{numericmotionstep}
Calculating the position of the $i^\mathrm{th}$ magnetic moment
in discrete time propelled by thermal energy:
\begin{equation}
x_i(k+1)=x_i(k)+s\,\eta(k),
\label{numericmotion}
\end{equation}
where $\eta(\cdot) \in \Rset^3$ is (pseudorandom) normally distributed.
\item
If the line through $(x_i(k+1),x_i(k))$ crosses a wall,
i.e.\ when hitting a wall,
correcting the path by
elastic collision computations,
\item
\label{numericdisplaceintegralstep}
Evaluating numerically the displacement integral differences of
Eq.~(\ref{stochasticphase1}) using
the displacements, $w_i(k):$
\begin{equation}
W^\mathrm{d}_i=t_{\mathrm{step}}\left(
\left(\sum\limits_{k=K_3}^{K_4}\,w_i(k)\right)-\left(\sum\limits_{k=K_1}^{K_2}\,w_i(k)\right)
\right)
\label{numericalW}
\end{equation}
where
$t_{\mathrm{step}} \times \left[\matrix{K_1 & K_2 & K_3 & K_4}\right]=
\left[\matrix{t_{d1} & t_{d2} & t_{d3} & t_{d4}}\right].$

\item Repeating
Step~\ref{numericmotionstep} through Step~\ref{numericdisplaceintegralstep}
for all of the magnetic moments and

\item Computing the distribution, $P_{\mathrm{cfd}}^{\mathrm{nmr}},$
of the displacement integral values.
\end{enumerate}

In--house Matlab\textregistered{}, (ver.\ 2014b, Mathworks, Natick, MA USA)
programs were used for all of the computations and visualization.
Distributions with $128$ bins were calculated using {\sf hist} and {\sf hist3} functions
for one and two dimensional simulations respectively.
For implementing discrete Fourier transform (DFT)
Matlab\textregistered{}'s
fast Fourier transform (FFT) routines
were used.

\section{Results and Discussion}
\begin{figure}[h!]
\centering
\includegraphics[width=12cm]{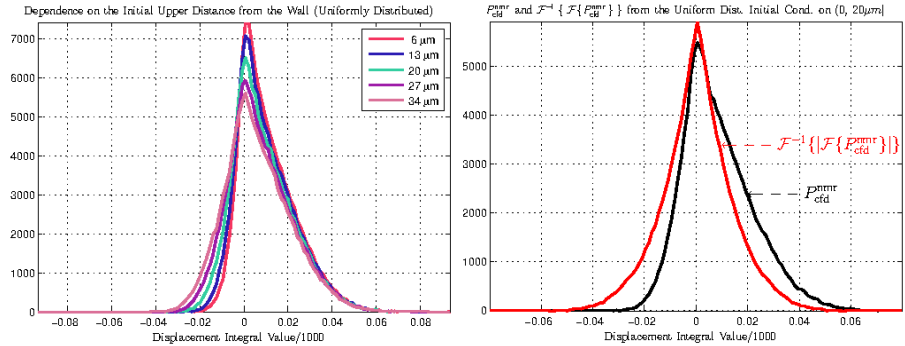}
\caption{On the left, the displacement integral distributions corresponding to
one dimensional simulations with a reflective wall placed at the origin.
The legend indicates color coding of the curves depending on
the upper limit of the initial condition interval.
As the upper limit
increases, i.e.\ more particles start far away from the wall,
$P_{\mathrm{cfd}}^{\mathrm{nmr}}$ becomes more symmetric and spread,
appearing more like isotropic motion.
On the right, the mismatch between the original
{\it asymmetric} distribution (black)
and the Fourier inversion of the magnitude (red) of the
distribution's Fourier transform,
$\mathcal F^{-1} \{| \mathcal{F}\{P_{\mathrm{cfd}}^{\mathrm{nmr}}\}|\},$
is shown for the initial condition interval $(0\; 20\mu m].$
Note the larger number of zero displacement integrals that are
inaccurately inferred using the distribution (red) obtained with signal magnitude.
}
\label{initialdistancewallfig}
\end{figure}
Although there are solutions in
simple microstructures with planar, spherical and
cylindrical geometries~\cite{Cotts1966264,Robertson1966273,BARZYKIN1999342,callaghan2011},
a full description of information loss from signal magnitude usage
in a general geometry is difficult to obtain
due to the lack of analytical characterizations of the distribution
functions.
Herein,
Brownian motion near reflecting walls in one and two dimensions,
similar to the
work in~\cite{ozarslan2009834,ozarslan20082809}, was simulated
with the algorithm of Section~\ref{simulationssection}
for numerically calculating the distributions.

For one dimensional scenario, a wall was placed on the origin and
particles were initially uniformly distributed
in an interval starting from the origin ending at $5$ different positions:
$(0\; 6\mu m],\, (0 \;13 \mu m], \, (0 \;20\mu m],\, (0 \;27\mu m],\, (0 \;34\mu m]$
(see Fig.~\ref{initialdistancewallfig}).
Simulations
were executed and $P_{\mathrm{cfd}}^{\mathrm{nmr}}(W^\mathrm{d})$
was obtained from the numerically calculated $W^\mathrm{d}_i,\:  i=1,\ldots,n_\mathrm{particles}.$

Figure~\ref{initialdistancewallfig} compares distribution functions
obtained from different initial conditions.
Initial condition intervals extending farther from the wall
resulted in more symmetric and spread $P_{\mathrm{cfd}}^{\mathrm{nmr}},$
resembling more isotropic motion.
In contrast, $P_{\mathrm{cfd}}^{\mathrm{nmr}}$'s {\it asymmetry}
is more prominent when particles start near the wall.

For all of the distributions, $\mathcal{F}\{P_{\mathrm{cfd}}^{\mathrm{nmr}}\}$
was calculated for simulating DW--MR signal coming out of the MR scanner.
As all of the distributions were {\bf asymmetric}, their Fourier transforms
were {\bf complex valued Hermitian}.
Therefore, the magnitude, $| \mathcal{F}\{P_{\mathrm{cfd}}^{\mathrm{nmr}}\}|,$
was a real valued symmetric function.
As the (inverse) Fourier transform of a real symmetric function is also real
symmetric~\cite{bracewell}, $\mathcal F^{-1} \{| \mathcal{F}\{P_{\mathrm{cfd}}^{\mathrm{nmr}}\}|\}$
was real and symmetric.
This is demonstrated on the right of Fig.~\ref{initialdistancewallfig} by
showing the mismatch between $P_{\mathrm{cfd}}^{\mathrm{nmr}}$ (asymmetric) and
$\mathcal F^{-1} \{| \mathcal{F}\{P_{\mathrm{cfd}}^{\mathrm{nmr}}\}|\}$
(symmetric).

In the one dimensional scenario which did not incorporate any bulk motion,
using signal magnitude,
which was claimed to be filtering out bulk motion~\cite[pp.1378]{wedeen05},
should have left the distribution intact.
Figure~\ref{initialdistancewallfig} shows that distribution asymmetry
is lost and the symmetric distribution
obtained from signal magnitude implies wrongly an isotropic medium.
In reality, for particles near an elastic wall the motion is anisotropic
as reported by the asymmetry of the original distribution.

\begin{figure}[h!]
\centering
\includegraphics[width=12cm]{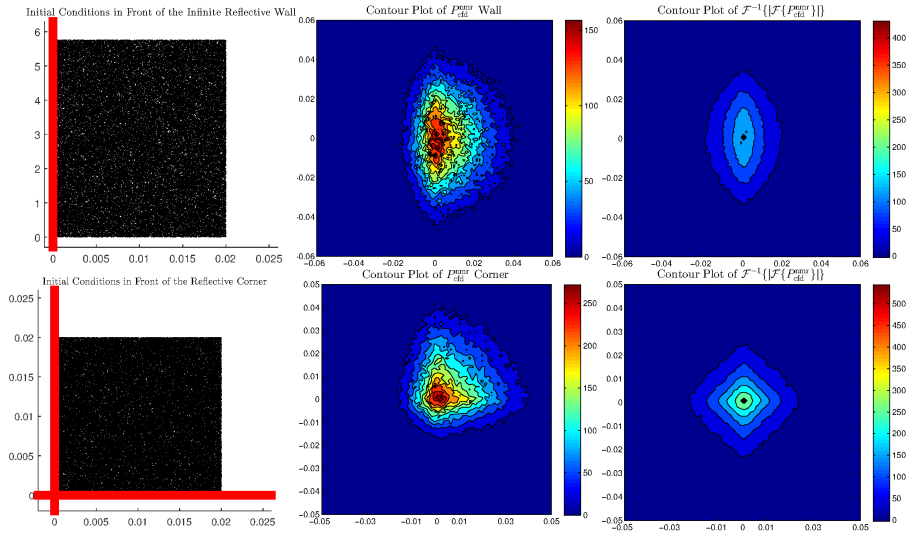}
\caption{On the left, initial uniform particle placement for the infinite reflective wall (top)
and the corner (bottom) scenarios.
The contour plots of the displacement integral distributions are shown
in the middle panel
demonstrating accurately the trend of particles bouncing from the wall(s).
On the top,
the distribution is skewed to the right implying the presence of the wall.
On the bottom, spread beam shaped distribution indicates the presence of a corner.
On the right column, using signal magnitude
creates inaccurate distributions.
The wall is inferred as a vertically oriented tube (top) and the corner
appears as a box (bottom). Furthermore, there is a large spike at the
origin of both distributions inaccurately indicating a large number
of particles with zero displacement integrals.
}
\label{numericalPcfd2Dfig}
\end{figure}%
In two dimensions, two scenarios
were run using a vertical wall and a corner,
both reflective and placed at the origin (see Fig.~\ref{numericalPcfd2Dfig}).
For the vertical wall, the initial positions were uniformly
distributed on the $(0,\; 20\mu m] \times (0,\; 5.76 mm]$ strip
($5.76 mm=0.1 \times s \times n_\mathrm{particles}$)
and
for the corner, the initial distribution was uniform on
the square: $(0,\; 20\mu m] \times (0,\; 20\mu m]$ (Fig.~\ref{numericalPcfd2Dfig}, left panel).

In Fig.~\ref{numericalPcfd2Dfig}'s middle panel, accurate inference
of the wall geometry was attained as original distributions were analyzed.
By contrast, on the right panel,
the Fourier transformed magnitude produces symmetric distributions
with the appearance of a tube (top) and a box (bottom) rather than a vertical wall
and a corner respectively.

Also, taking the magnitude transfers the
imaginary portion to the real axis, thereby adding a positive
constant to the signal which in turn
becomes a spike at $[0,\, 0]$ in Fig.~\ref{numericalPcfd2Dfig}
and at $0$ in Fig.~\ref{initialdistancewallfig}
when calculating the Fourier transform.
This, in turn, inaccurately indicates a large number of
magnetic moments with zero
displacement integral values,
which might mislead inference of mathematical models
that include terms for stationary magnetic moments, e.g.,
of water trapped in
small compartments~\cite{ALEXANDER20101374,Panagiotaki20122241},
likewise the positive constant term in models using cumulant
expansion~\cite{kroenke06}.

\section{Conclusions}
Biological interpretation of DW--MR information
is the subject of active research with new developments
using existing model attributes, such as the decrease in the
apparent diffusion coefficient caused by neurite beading
after ischemic stroke~\cite{Budde10082010} and cerebral
cortical gray matter development~\cite{kroenke2018106}.
New theories describing novel attributes are also being
developed for the same purpose, e.g.,\ the MAP model~\cite{ozarslan201316}
where the focus is the microstructure in brain tissue.

The purpose of the DW--MR, like any other imaging modality,
is the inference of biological properties with the aim
of producing informative biomarkers.
This is accomplished in two steps starting from the MR scanner's output signal
that is Fourier transformed for reconstructing the physically meaningful
displacement integral distribution:
\[
S_{\mathrm{cfd}} \rightarrow P_{\mathrm{cfd}} \rightarrow \mbox{Biological properties}.
\]
In the process, intermediate steps must be carried out carefully
while respecting physically meaningful information content.

This work demonstrated that, as shown in Fig.~\ref{initialdistancewallfig} and Fig.~\ref{numericalPcfd2Dfig},
the usage of magnitude may result in incorrect information:
\begin{equation}
\mathcal F^{-1} \{|S_{\mathrm{cfd}}^{\mathrm{nmr}}|\}
\neq P_{\mathrm{cfd}}^{\mathrm{nmr}}
=\mathcal F^{-1} \{S_{\mathrm{cfd}}^{\mathrm{nmr}}\},
\end{equation}
potentially hampering accurate inference of microstructure.

The proof was achieved in an ideal simulated environment
with minimal assumptions, e.g.\ without including loss of magnetization
at the walls~\cite{SNAAR1993318,Sen1994215},
wall permeability~\cite{Tanner19781748} and particle interactions.
While no mathematical model is perfect, falsification when validating under
ideal conditions is a legitimate reason for abandoning a model
or at least using prudence in interpreting the model's outcomes.
For future in and ex vivo
diffusion imaging research,
rather than assuming that diffusion profiles in biological tissue
are symmetric for any reason, it would be prudent to adopt DW--MR models
incorporating asymmetry assumptions and let the data analysis
indicate symmetry and/or lack thereof.

Accordingly, instead of using signal magnitude,
the solution offered by CFD--MR~\cite{alpaycfd} is based
on re--establishing Hermitian symmetry of the complex
valued signal collected from the MR scanner by using systemic phase correction
algorithms. For example, bulk motion appears as a linear phase
which is rigorously handled with a linear phase correction algorithm
in the Fourier domain~\cite{alpaycfd}.
However, factors which are beyond the scope of this manuscript,
including but not limited to susceptibility effects,
distort the Hermitian symmetry of the DW--MR signal
in a nonlinear fashion. Development of algorithms for
analyzing and subsequently correcting those effects remains
an open question for future research.
Furthermore, using complex valued signal in CFD--MR automatically
removes nonlinear Rician restrictions~\cite{gudbjartsson} on noise modeling
in DW--MR.

In conclusion, this paper mainly concentrated
on the first step that constitutes the foundation for biological property inference,
namely removing signal magnitude processing
from DW--MR methods with the aim of improving accurate reconstruction
of physical quantities.
Simulations in fundamental geometries demonstrated that
the usage of signal magnitude in diffusion MR can significantly
distort the estimation of motion describing distributions,
and hence may impede accurate inference of microstructure.

\section*{Acknowledgements}
This work was supported by the Scientific and Technological Research Council
of Turkey's (TUBITAK) $3001$ funding program,
titled ``Simulating Microstructure Inferred from Diffusion
Magnetic Resonance Imaging and Modelling Effects of
the MR Scanner on the Diffusion MR Signal'', grant number $118$E$304.$

The author would also like to extend his special thanks to
John Gore and the reviewers for their valuable insights
and progressive feedback.


\begin{thebibliography}{10}
\expandafter\ifx\csname url\endcsname\relax
  \def\url#1{\texttt{#1}}\fi
\expandafter\ifx\csname urlprefix\endcsname\relax\def\urlprefix{URL }\fi
\expandafter\ifx\csname href\endcsname\relax
  \def\href#1#2{#2} \def\path#1{#1}\fi

\bibitem{alpaybackground12}
A.~{\"O}zcan, K.~H. Wong, L.~Larson-Prior, Z.-H. Cho, S.~K. Mun,
  \href{http://dx.doi.org/10.1002/ima.22001}{Background and mathematical
  analysis of diffusion {MRI} methods}, International Journal of Imaging
  Systems and Technology 22~(1) (2012) 44--52.
\newblock \href {https://doi.org/10.1002/ima.22001}
  {\path{doi:10.1002/ima.22001}}.
\newline\urlprefix\url{http://dx.doi.org/10.1002/ima.22001}

\bibitem{ozarslan201316}
E.~{\"O}zarslan, C.~G. Koay, T.~M. Shepherd, M.~E. Komlosh, M.~O.
  {\.I}rfano{\=g}lu, C.~Pierpaoli, P.~J. Basser,
  \href{http://www.sciencedirect.com/science/article/pii/S1053811913003431}{Mean
  apparent propagator ({MAP}) {MRI}: A novel diffusion imaging method for
  mapping tissue microstructure}, NeuroImage 78~(0) (2013) 16 -- 32.
\newblock \href {https://doi.org/10.1016/j.neuroimage.2013.04.016}
  {\path{doi:10.1016/j.neuroimage.2013.04.016}}.
\newline\urlprefix\url{http://www.sciencedirect.com/science/article/pii/S1053811913003431}

\bibitem{moseley04gdt}
C.~Liu, R.~Bammer, B.~Acar, M.~E. Moseley, Characterizing non--gaussian
  diffusion by using generalized diffusion tensors, Magnetic Resonance in
  Medicine 51~(5) (2004) 924--937.

\bibitem{alpaycfd}
A.~{\"O}zcan,
  \href{http://www.frontiersin.org/integrative{\_}neuroscience/10.3389/fnint.2013.00018/abstract}{Complete
  fourier direct magnetic resonance imaging ({CFD}--{MRI}) for diffusion
  {MRI}}, Frontiers in Integrative Neuroscience 7~(18) (2013).
\newblock \href {https://doi.org/10.3389/fnint.2013.00018}
  {\path{doi:10.3389/fnint.2013.00018}}.
\newline\urlprefix\url{http://www.frontiersin.org/integrative{\_}neuroscience/10.3389/fnint.2013.00018/abstract}

\bibitem{basser94}
P.~Basser, J.~Mattiello, D.~Lebihan,
  \href{http://www.sciencedirect.com/science/article/pii/S1064186684710375}{Estimation
  of the effective self--diffusion tensor from the {NMR} spin echo}, Journal of
  Magnetic Resonance, Series B 103~(3) (1994) 247 -- 254.
\newblock \href {https://doi.org/10.1006/jmrb.1994.1037}
  {\path{doi:10.1006/jmrb.1994.1037}}.
\newline\urlprefix\url{http://www.sciencedirect.com/science/article/pii/S1064186684710375}

\bibitem{matiello94}
J.~Mattiello, P.~Basser, D.~Lebihan,
  \href{http://www.sciencedirect.com/science/article/pii/S106418588471103X}{Analytical
  expressions for the b matrix in {NMR} diffusion imaging and spectroscopy},
  Journal of Magnetic Resonance, Series A 108~(2) (1994) 131 -- 141.
\newblock \href {https://doi.org/10.1006/jmra.1994.1103}
  {\path{doi:10.1006/jmra.1994.1103}}.
\newline\urlprefix\url{http://www.sciencedirect.com/science/article/pii/S106418588471103X}

\bibitem{frank01}
L.~R. Frank, Anisotropy in high angular resolution diffusion--weighted {MRI},
  Magnetic Resonance in Medicine 45~(6) (2001) 935--1141.

\bibitem{callaghan2011}
P.~T. Callaghan, Translational Dynamics and Magnetic Resonance: Principles of
  Pulsed Gradient Spin Echo {NMR}, Oxford University Press, Oxford [England]
  New York, 2011.

\bibitem{wedeen05}
V.~J. Wedeen, P.~Hagmann, W.-Y.~I. Tseng, T.~G. Reese, R.~M. Weisskoff, Mapping
  complex tissue architecture with diffusion spectrum magnetic resonance
  imaging, Magnetic Resonance in Medicine 54~(6) (2005) 1377--1385.

\bibitem{tuch04}
D.~S. Tuch, Q--ball imaging, Magnetic Resonance in Medicine 52~(6) (2004)
  1358--1372.

\bibitem{Cotts1966264}
R.~C. Wayne, R.~M. Cotts,
  \href{https://link.aps.org/doi/10.1103/PhysRev.151.264}{Nuclear--magnetic--resonance
  study of self--diffusion in a bounded medium}, Phys. Rev. 151 (1966)
  264--272.
\newblock \href {https://doi.org/10.1103/PhysRev.151.264}
  {\path{doi:10.1103/PhysRev.151.264}}.
\newline\urlprefix\url{https://link.aps.org/doi/10.1103/PhysRev.151.264}

\bibitem{Robertson1966273}
B.~Robertson,
  \href{https://link.aps.org/doi/10.1103/PhysRev.151.273}{Spin--echo decay of
  spins diffusing in a bounded region}, Phys. Rev. 151 (1966) 273--277.
\newblock \href {https://doi.org/10.1103/PhysRev.151.273}
  {\path{doi:10.1103/PhysRev.151.273}}.
\newline\urlprefix\url{https://link.aps.org/doi/10.1103/PhysRev.151.273}

\bibitem{BARZYKIN1999342}
A.~Barzykin,
  \href{http://www.sciencedirect.com/science/article/pii/S1090780799917780}{Theory
  of spin echo in restricted geometries under a step--wise gradient pulse
  sequence}, Journal of Magnetic Resonance 139~(2) (1999) 342 -- 353.
\newblock \href {https://doi.org/https://doi.org/10.1006/jmre.1999.1778}
  {\path{doi:https://doi.org/10.1006/jmre.1999.1778}}.
\newline\urlprefix\url{http://www.sciencedirect.com/science/article/pii/S1090780799917780}

\bibitem{haacke}
R.~Brown, Y.~Cheng, E.~Haacke, M.~Thompson, R.~Venkatesan, Magnetic Resonance
  Imaging: Physical Principles and Sequence Design, John Wiley and Sons, New
  York, 2014.

\bibitem{alpayisbi11}
A.~{\"O}zcan, Comparison of the {Complete} {Fourier} {Direct} {MRI} with
  existing diffusion weighted {MRI} methods, in: Proceedings of the 2011 IEEE
  International Symposium on Biomedical Imaging, Chicago, Illinois, USA, 2011,
  pp. 931--934.

\bibitem{bracewell}
R.~N. Bracewell, The Fourier transform and its applications, 3rd Edition,
  McGraw-Hill series in electrical and computer engineering. Circuits and
  systems, McGraw Hill, Boston, 2000, (DLC) 99021139 Ronald N. Bracewell. ill.
  ; 25 cm. Includes bibliographical references and index.

\bibitem{Ermak19754189}
D.~L. Ermak, \href{https://doi.org/10.1063/1.430300}{A computer simulation of
  charged particles in solution. {I}. {Technique} and equilibrium properties},
  The Journal of Chemical Physics 62~(10) (1975) 4189--4196.
\newblock \href {https://doi.org/10.1063/1.430300}
  {\path{doi:10.1063/1.430300}}.
\newline\urlprefix\url{https://doi.org/10.1063/1.430300}

\bibitem{alpayembc11}
A.~{\"O}zcan, J.~Quirk, Y.~Wang, Q.~Wang, P.~Sun, W.~Spees, S.-K. Song, The
  validation of complete fourier direct {MR} method for diffusion {MRI} via
  biological and numerical phantoms, in: Engineering in Medicine and Biology
  Society,EMBC, 2011 Annual International Conference of the IEEE, 2011, pp.
  3756 --3759.
\newblock \href {https://doi.org/10.1109/IEMBS.2011.6090640}
  {\path{doi:10.1109/IEMBS.2011.6090640}}.

\bibitem{ozarslan2009834}
E.~{\"O}zarslan, C.~G. Koay, P.~J. Basser,
  \href{http://www.sciencedirect.com/science/article/pii/S0730725X09000113}{Remarks
  on q--space {MR} propagator in partially restricted, axially--symmetric, and
  isotropic environments}, Magnetic Resonance Imaging 27~(6) (2009) 834 -- 844.
\newblock \href {https://doi.org/10.1016/j.mri.2009.01.005}
  {\path{doi:10.1016/j.mri.2009.01.005}}.
\newline\urlprefix\url{http://www.sciencedirect.com/science/article/pii/S0730725X09000113}

\bibitem{ozarslan20082809}
E.~{\"O}zarslan, U.~Nevo, P.~J. Basser,
  \href{http://www.sciencedirect.com/science/article/pii/S0006349508705320}{Anisotropy
  induced by macroscopic boundaries: Surface--normal mapping using
  diffusion--weighted imaging}, Biophysical Journal 94~(7) (2008) 2809 -- 2818.
\newblock \href {https://doi.org/10.1529/biophysj.107.124081}
  {\path{doi:10.1529/biophysj.107.124081}}.
\newline\urlprefix\url{http://www.sciencedirect.com/science/article/pii/S0006349508705320}

\bibitem{ALEXANDER20101374}
D.~C. Alexander, P.~L. Hubbard, M.~G. Hall, E.~A. Moore, M.~Ptito, G.~J.
  Parker, T.~B. Dyrby,
  \href{http://www.sciencedirect.com/science/article/pii/S1053811910007755}{Orientationally
  invariant indices of axon diameter and density from diffusion {MRI}},
  NeuroImage 52~(4) (2010) 1374 -- 1389.
\newblock \href
  {https://doi.org/https://doi.org/10.1016/j.neuroimage.2010.05.043}
  {\path{doi:https://doi.org/10.1016/j.neuroimage.2010.05.043}}.
\newline\urlprefix\url{http://www.sciencedirect.com/science/article/pii/S1053811910007755}

\bibitem{Panagiotaki20122241}
E.~Panagiotaki, T.~Schneider, B.~Siow, M.~G. Hall, M.~F. Lythgoe, D.~C.
  Alexander,
  \href{http://www.sciencedirect.com/science/article/pii/S1053811911011566}{Compartment
  models of the diffusion {MR} signal in brain white matter: A taxonomy and
  comparison}, NeuroImage 59~(3) (2012) 2241 -- 2254.
\newblock \href
  {https://doi.org/http://dx.doi.org/10.1016/j.neuroimage.2011.09.081}
  {\path{doi:http://dx.doi.org/10.1016/j.neuroimage.2011.09.081}}.
\newline\urlprefix\url{http://www.sciencedirect.com/science/article/pii/S1053811911011566}

\bibitem{kroenke06}
C.~D. Kroenke, G.~L. Bretthorst, T.~E. Inder, J.~J. Neil, Modeling water
  diffusion anisotropy within fixed newborn primate brain using bayesian
  probability theory, Magnetic Resonance in Medicine 55~(1) (2006) 187--197.

\bibitem{Budde10082010}
M.~D. Budde, J.~A. Frank,
  \href{http://www.pnas.org/content/107/32/14472.abstract}{Neurite beading is
  sufficient to decrease the apparent diffusion coefficient after ischemic
  stroke}, Proceedings of the National Academy of Sciences 107~(32) (2010)
  14472--14477.
\newblock \href {https://doi.org/10.1073/pnas.1004841107}
  {\path{doi:10.1073/pnas.1004841107}}.
\newline\urlprefix\url{http://www.pnas.org/content/107/32/14472.abstract}

\bibitem{kroenke2018106}
C.~D. Kroenke,
  \href{http://www.sciencedirect.com/science/article/pii/S1090780718301150}{Using
  diffusion anisotropy to study cerebral cortical gray matter development},
  Journal of Magnetic Resonance 292 (2018) 106 -- 116.
\newblock \href {https://doi.org/https://doi.org/10.1016/j.jmr.2018.04.011}
  {\path{doi:https://doi.org/10.1016/j.jmr.2018.04.011}}.
\newline\urlprefix\url{http://www.sciencedirect.com/science/article/pii/S1090780718301150}

\bibitem{SNAAR1993318}
J.~Snaar, H.~Vanas,
  \href{http://www.sciencedirect.com/science/article/pii/S1064185883711101}{{NMR}
  self-diffusion measurements in a bounded system with loss of magnetization at
  the walls}, Journal of Magnetic Resonance, Series A 102~(3) (1993) 318 --
  326.
\newblock \href {https://doi.org/https://doi.org/10.1006/jmra.1993.1110}
  {\path{doi:https://doi.org/10.1006/jmra.1993.1110}}.
\newline\urlprefix\url{http://www.sciencedirect.com/science/article/pii/S1064185883711101}

\bibitem{Sen1994215}
P.~N. Sen, L.~M. Schwartz, P.~P. Mitra, B.~I. Halperin,
  \href{https://link.aps.org/doi/10.1103/PhysRevB.49.215}{Surface relaxation
  and the long--time diffusion coefficient in porous media: Periodic
  geometries}, Phys. Rev. B 49 (1994) 215--225.
\newblock \href {https://doi.org/10.1103/PhysRevB.49.215}
  {\path{doi:10.1103/PhysRevB.49.215}}.
\newline\urlprefix\url{https://link.aps.org/doi/10.1103/PhysRevB.49.215}

\bibitem{Tanner19781748}
J.~E. Tanner, \href{https://doi.org/10.1063/1.436751}{Transient diffusion in a
  system partitioned by permeable barriers. {Application} to {NMR} measurements
  with a pulsed field gradient}, The Journal of Chemical Physics 69~(4) (1978)
  1748--1754.
\newblock \href {https://doi.org/10.1063/1.436751}
  {\path{doi:10.1063/1.436751}}.
\newline\urlprefix\url{https://doi.org/10.1063/1.436751}

\bibitem{gudbjartsson}
H.~Gudbjartsson, S.~Patz, The {Rician} distribution of noisy {MRI} data,
  Magnetic Resonance in Medicine 34~(6) (1995) 910--914.

\end{thebibliography}
\end{document}